\documentclass[useAMS,usenatbib,usegraphicx]{mn2e}

\usepackage{graphicx}
\usepackage{subfigure}
\usepackage{times,journals}

\usepackage{color}

\definecolor{grey}{rgb}{0.5,0.6,0.7}
\def \simlt { \lower .75ex \hbox{$\sim$} \llap{\raise .27ex \hbox{$<$}} }
\definecolor{purple}{rgb}{0.65,0.15,0.9}
\definecolor{darkorange}{rgb}{0.8,0.3,0}
\definecolor{olive}{rgb}{0.4,0.6,0.25}
\definecolor{darkgreen}{rgb}{0,0.7,0}
\definecolor{darkred}{rgb}{0.5,0,0}

\title{Testing primordial non-Gaussianities on galactic scales at high
  redshift}

\author[Habouzit et al.]{M\'elanie Habouzit$^1$\thanks{E-mail: habouzit@iap.fr}, 
Takahiro~Nishimichi$^1$,
S\'ebastien~Peirani$^1$,
Gary A. Mamon$^1$, 
\newauthor
Joseph Silk$^{1,2}$,
and Jacopo Chevallard$^1$\\
$^1$Institut d'Astrophysique de Paris (UMR 7095: CNRS \& UPMC), 98 bis
Bd Arago, F--75014 Paris, France\\
$^2$Department of Physics and Astronomy, The Johns Hopkins University
Homewood Campus, Baltimore MD 21218, USA}

\begin{document}

\date{Accepted yyyy month dd. Received yyyy month dd; in original form yyyy
  month dd} 


\maketitle

\label{firstpage}

\begin{abstract}
Primordial non-Gaussianities  
provide an important test of inflationary models. Although the \emph{Planck} CMB experiment has produced strong limits on
non-Gaussianity on scales of clusters, there is still room for considerable non-Gaussianity on galactic scales.
We have tested the effect of local non-Gaussianity on the high redshift galaxy population by running five cosmological $N$-body simulations down to $z=6.5$. For these simulations, we adopt the same initial phases, and either Gaussian or scale-dependent non-Gaussian primordial fluctuations, all consistent with the constraints set by \emph{Planck} on cluster scales.
We then assign stellar masses to each halo using the halo -- stellar mass empirical relation of Behroozi et al. (2013). Our simulations with non-Gaussian initial conditions produce halo mass functions that show clear departures from those obtained from the analogous simulations with Gaussian initial conditions at $z \ga 10$. We observe a $>0.3$ dex enhancement of the low-end of the halo mass function, which leads to a similar effect on the galaxy stellar mass function, which should be testable with future galaxy surveys at $z>10$.
As cosmic reionization is thought to be driven by dwarf galaxies at high redshift, our findings may have implications for the reionization history of the Universe.
\end{abstract}


\section{Introduction}

The simplest inflationary models predict a very nearly Gaussian distribution of density 
perturbations \citep{Gangui+94,Acquaviva2002,Maldacena03}.
Primordial non-Gaussianities are therefore an important test of how physics shaped the
universe at early times, at energies too high to be probed by laboratory experiments. 
The departures from Gaussianity at the leading order are characterized by the \emph{bispectrum},
the Fourier counterpart of the 3-point correlation function, and models are often classified
by the triangular configuration of wavevectors at which the bispectrum has the largest signal.

Most common is the \emph{local}-type non-Gaussianity,
for which the bispectrum is maximal when two of the wavenumbers are much greater than 
the third one \citep{Gangui+94}. 
The magnitude of the non-Gaussianity of this type can be parameterized \citep{Komatsu+01} by a
parameter, $f_{\rm NL}$, describing 
the quadratic coupling of the primordial perturbations
\begin{eqnarray}
\zeta(\mathbf{x})=
\zeta_\rmn{G}(\mathbf{x})+\frac{3}{5}f_{\rm NL} \left(\zeta_\rmn{G}^{2}(\mathbf{x})-\left\langle
\zeta_\rmn{G}^{2}(\mathbf{x})\right\rangle \right) \ ,
\end{eqnarray}
where $\zeta$ is the curvature perturbations and $\zeta_\rmn{G}$ is a Gaussian random field at the same position.
Standard single-field inflationary theories predict
$f_\rmn{NL} \sim \varepsilon$, where
$\varepsilon \ll 1$ is the slow roll parameter, which is independent
of scale \citep{Maldacena03}.
The primordial density fluctuations evolve with time, and lead to the collapse of dark matter particles, and baryons. 
A non-Gaussian initial spectrum of density perturbations will then affect the distribution of baryonic structures.
The \emph{Planck} mission, which mapped in detail the Cosmic Microwave
Background (CMB) on the full sky,
has provided a
much stronger constraint on the local non-Gaussianity parameter,
$f_{\rm NL} = 2.7 \pm 5.8$
\citep{PlanckCollaboration+13}, than did the previous CMB mission (the Wilkinson Microwave Anisotropy
Probe, \emph{WMAP}, \citealp{Bennett+13}).

By appealing to the theory of \cite{Press74} it is straightforward to show that positively skewed ($f_{\rm NL} > 0$) primordial density
fluctuations increase the halo mass function (HMF) at large masses with respect to that arising from Gaussian initial conditions (e.g. \citealp{Matarrese:2000lr}). This effect has also been checked with cosmological $N$-body simulations \citep{Kang+07,Grossi+07,Pillepich+10}.
Simulations with non-Gaussian initial conditions (nGICs) have been used to probe
 the halo mass function
(\citealp{Kang+07}; \citealp{Grossi+07}; \citealp*{Pillepich+10}), the
scale-dependent halo bias (\citealp{Dalal+08}; \citealp{Desjaques+09};
\citealp{Grossi+09}) and
bispectrum (\citealp{Nishimichi+10}; \citealp*{Sefusatti+10}), 
weak lensing statistics (\citealp{Pace+11}; \citealp*{Shirasaki+12}),
the pairwise velocity distribution
function \citep*{Lam+11}. Hydrodynamical cosmological simulations have been
performed with nGICs to study the baryon history
\citep{Maio+11}, the gas distribution \citep{Maio11}, the gas density
profiles \citep{Maio+12} and SZ maps \citep{Pace+14}.

These studies have all used a scale-independent value of $f_{\rm NL}$. 
However, the new constraint on $f_{\rm NL}$ on large scales does not exclude non-Gaussianity on smaller scales, namely galactic scales.  
Indeed, the non-Gaussianity might depend on scale, as predicted, e.g., in
several inflation models with a variable speed of sound, such as the
string-based Dirac-Born-Infeld models \citep*{Silverstein+04,Alishahiha+04,Chen:2005fk}.

It is thus possible that significant non-Gaussianity can lurk on the comoving
scales of galaxies without being detected by the \emph{Planck} CMB mission, whose
angular resolution effectively limits it to the scales of clusters of
galaxies. 
A \emph{blue} spectrum of \emph{running non-Gaussianity}
might enhance low masses instead, if the spectrum is blue
enough, i.e. if  ${\rm d}\ln f_{\rm NL} / {\rm d}\ln k$ is large enough.
The effects of scale-dependent non-Gaussianity on the HMF (cluster counts)
and on reionization
were analytically predicted by \cite{LoVerde+08} and \cite{Crociani2009}, respectively.
Because small scales are still poorly constrained at high redshifts ($z>6$),
cosmological simulations are key for
predicting whether non-Gaussianities have an impact on galactic scales.
Only one team has run nG simulations with an {\it explicit scale-dependence} adjustable by
a free parameter \citep{Shandera+11}, focusing on halo clustering in the the local Universe.

The present work aims to predict, analytically and with cosmological $N$-body
simulations, 
the effects of running non-Gaussianity on the galaxy
stellar mass function (SMF), focusing on high redshifts ($z>6$, i.e. less than 950 Myr after the Big Bang), 
where the effects of primordial
non-Gaussianity ought to be most important, and on masses low enough that one
may see the reverse of the enhancement of the SMF caused by
$f_{\rm NL}>0$, from the high end to the low end. 
Future galaxy surveys with \emph{Euclid} or the \emph{James Webb Space
  Telescope} may soon probe these fairly low masses at very high redshifts.

The paper is organized as follows. In Section 2, we present our simulations, in particular the set-up of the nGICs, as well as our adopted galaxy
formation and evolution model. 
Section 3 begins with an analytical prediction of the HMF arising from nGICs, and then 
we compare both HMFs and SMFs derived from our simulations with nGICs  with those from our Gaussian simulations.
Finally, we summarize and discuss our results in Section 4.

\vspace {-\baselineskip}

\section{Methodology}
\label{sec:methodology}

\subsection{Initial conditions: prescription for $f_{\rm NL}(k)$}
\label{subsec:fnl}
We employed a simple model that allows a significant amount of
non-Gaussianity on small scales, relevant for early structure formation,
while keeping such effects  small on large scales to meet the strong constraints obtained by the
\emph{Planck} CMB mission \citep{PlanckCollaboration+13}.
Namely, we investigated here the {\it generalized local ansatz} proposed
by \cite{Becker+11}:
\begin{equation} 
\zeta(\mathbf{x}) = \zeta_\rmn{G}(\mathbf{x}) + \frac{3}{5}\left[f_\rmn{NL} \star
  (\zeta_\rmn{G}^2 - \langle{\zeta_\rmn{G}^2}\rangle)\right](\mathbf{x}) \ ,
\label{eq:localansatz}
\end{equation}
where the operation $(f_\rmn{NL} \star A)$ is the convolution of a random variable $A$ and 
a $k$-dependent kernel defined in Fourier space:
\begin{equation} 
f_{\rmn{NL}}(k) = f_{\rmn{NL}, 0}\,\left(\frac{k}{k_0}\right)^{\alpha}.
\label{eq:nGmodel}
\end{equation}

\begin{figure}
\centering
\includegraphics[width=0.7\hsize]{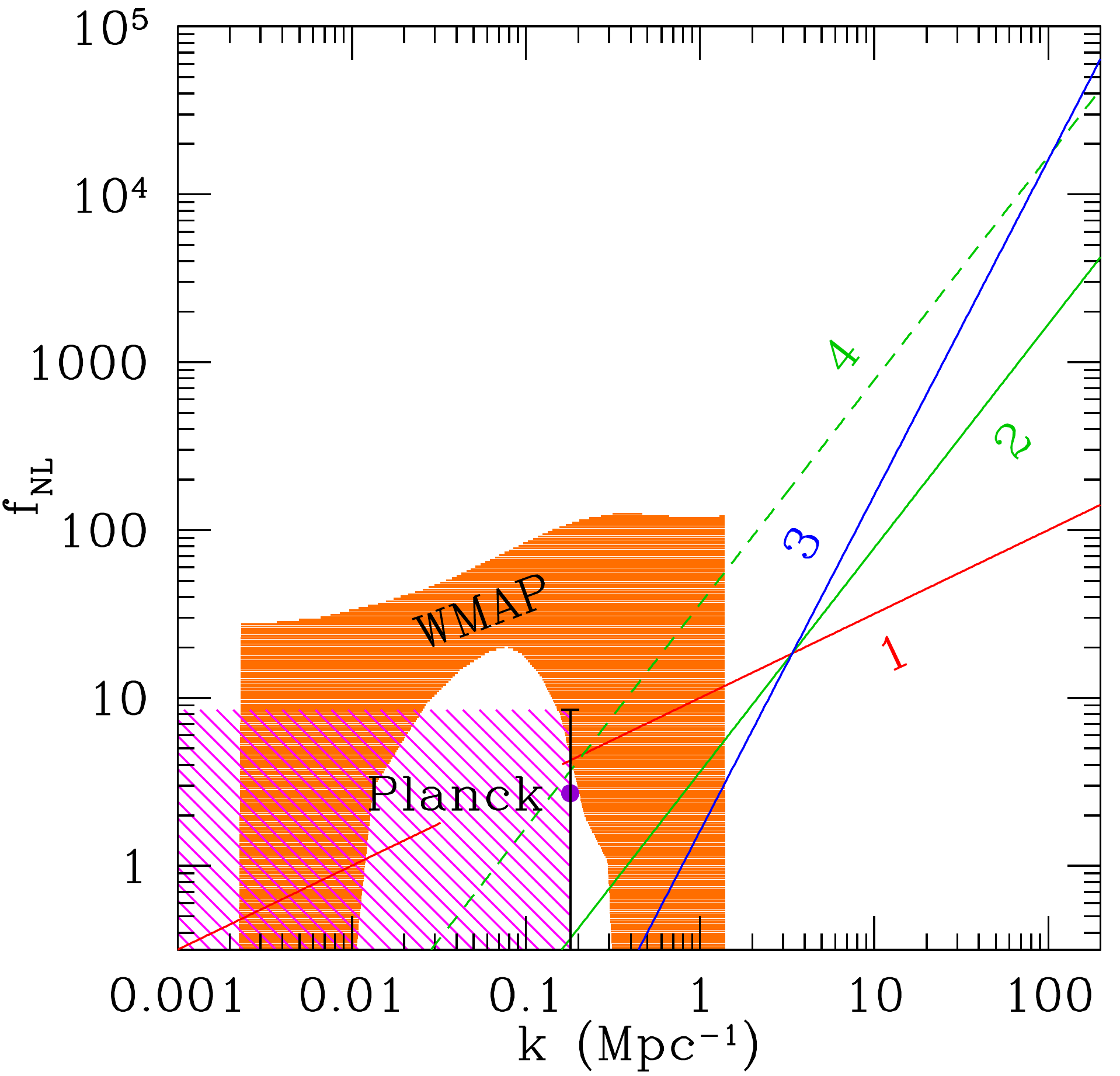} 
\caption{Models (\emph{lines}) for the scale-dependent non-Gaussian parameter 
$f_{\rm NL}(k)$ (eq.~[\ref{eq:nGmodel}], with parameters listed in 
Table~\ref{tab:fnlofk}).
The \emph{orange shaded region} represents the allowed values from \emph{WMAP}, within $1\,\sigma$, 
of $f_{\rm NL}(k)$
according to Becker \& Huterer (2012).
The \emph{magenta shaded region} shows the \emph{Planck}
constraint (Planck Collaboration et al. 2013b).
The right edge of the box corresponds to a scale of $2\pi/k \simeq 30 \,\rm
kpc$, i.e. the scales of galaxies are to the right of the right edge of the box.
\label{fig:fnlofk}
}
\end{figure}

We explored four different non-Gaussian (nG) models by varying the normalization
$f_{\rm NL,0}$ and the slope $\alpha={\rm d}\ln f_{\rm NL} / {\rm d}\ln k$, in such a way that
the non-Gaussianity is significant on galactic scales, yet small enough to
meet the current constraints from \emph{Planck} \citep{PlanckCollaboration+13}.
Table~\ref{tab:fnlofk}  (normalization and slope for $k_0 = 100\,h/\rmn{Mpc}$) lists our adopted models, while
Fig.~\ref{fig:fnlofk}  displays these models with current constraints from
CMB experiments.
We restricted ourselves to positively skewed primordial density fluctuations,
i.e. $f_{\rm NL}>0$, hence $f_{\rm NL,0}>0$.

\begin{table}
\caption{Characteristics of $f_{\rm NL}$ models (eq. [\ref{eq:nGmodel}]) 
\label{tab:fnlofk}}
\begin{center}
\begin{tabular}{cccccc}
\hline
Model & G & NG1 & NG2 & NG3 & NG4 \\
\hline
$f_{\rm NL,0}$ & 0 & 82 & 1000 & 7357 & 10000 \\
$\alpha$ & -- & 1/2 & 4/3 & 2 & 4/3 \\
\hline
\end {tabular}
\end{center}
\end{table}

We modified
the initial condition generator originally developed by \citet{Nishimichi+09},
based on second-order Lagrangian perturbation theory
(e.g., \citealp{Scoccimarro98,Crocce+06}), parallelized by \citet{Valageas+11}
and with local-type non-Gaussianities implemented by \citet{Nishimichi12}.
We followed \citet{Becker+11} and realized the generalized local ansatz of
equation (\ref{eq:localansatz}) by taking
a convolution of the curvature squared and the $k$-dependent $f_\mathrm{NL}$
kernel in Fourier space.  We used the public Boltzmann code, {\sc camb}
\citep{Lewis+00} to compute the transfer function and multiply it to the
curvature perturbations to have the linear density fluctuations.

\subsection{N-body simulations and halo catalog}

We have performed five cosmological simulations with {\sc Gadget-2}
\citep{Springel05} for a $\Lambda$CDM
universe using \emph{Planck} parameters \citep{PlanckCollaboration+13_cosmopars}, namely
$\Omega_M=0.307$,  $\Omega_{\Lambda}=0.693$,  $h=0.678$
and $\sigma_8=0.829$.
Each simulation was performed in a periodic box of side $50\,h^{-1}$ Mpc
with $1024^3$ dark matter particles (e.g. with mass resolution
of $\sim 9.9 \times 10^6\,h^{-1}\,\rm M_\odot$).
One simulation (hereafter, `G') started with Gaussian ICs, while the other
four (hereafter, `NG')
began with nGICs (eqs.~[\ref{eq:localansatz}] and
[\ref{eq:nGmodel}], with parameters in Table~\ref{tab:fnlofk}),
with the same initial phases.
The simulations started at $z=200$ and ended at $z=6.5$.
In each case, the Plummer-equivalent force softening 
was set to 5\% of the mean inter-particle distance
($2.44\,h^{-1}\,\rm kpc$ in comoving units).

For each snapshot (taken every $\sim40\, \rmn{Myr}$), catalogues of
halos were prepared using {\sc AdaptaHOP} \citep*{Aubert+04},
which employs an SPH-like kernel to compute
densities at the location of each particle and partitions
the ensemble of particles into (sub)halos based on saddle points
in the density field.
Only halos or subhalos containing at least 20 particles  (e.g. $2.9\times
10^8 \rmn{M}_\odot$) were retained.
We then studied the individual evolution of (sub)halos, by
building halo  merger trees using  {\sc TreeMaker} \citep{Tweed+09},
which allowed us to accurately derive the mass evolution of each dark
matter (sub)halo.
This was the basis to compute the evolution of galaxy stellar masses, as we shall
see in Sect.~\ref{sec:galform}. 

\vspace {-\baselineskip}

\subsection{Galaxy formation and evolution model}
\label{sec:galform}
Galaxy stellar masses are `painted' on the halos and
subhalos using the \cite*{Behroozi+13} model that provides the galaxy mass $m$ as a
function of halo mass $M$ and redshift $z$.
We could have adopted a \emph{physical} model, such as
\cite{Cattaneo+11}.
We also considered using the \emph{empirical} model of \cite*{Mutch+13}. 
The former model is only constrained at $z=0$, while the latter extends to
$z=4$, which is still insufficient for our purposes.
We have thus preferred to adopt the empirical model of
\citeauthor{Behroozi+13}, whose parameters were fit to the galaxy stellar mass
functions, specific star formation rates and cosmic star formation rate,
from $z=0$ to $z=8$. In particular, the \citeauthor{Behroozi+13} model is
the only empirical model of galaxy mass vs. halo mass and redshift that
extends up to the redshift when reionization is thought to occur.
A weakness of our approach is that, for lack of a better simple model,  we
assume that the \citeauthor{Behroozi+13} model can be extrapolated beyond
$z=8$ to $z=17$.

The \citeauthor{Behroozi+13} model was calibrated with HMFs derived
from cosmological simulations with Gaussian ICs. One could argue that their
model cannot be applied to simulations with nGICs, without appropriate
corrections. Alternatively, one could adopt the \citeauthor{Behroozi+13} model as a
basis to which we can compare the effects of Gaussian vs non-Gaussian ICs,
and this is what we do here.

However, we slightly modify the \citeauthor{Behroozi+13} model, by preventing galaxy masses 
from decreasing in time. For quiescent (sub)halos, we simply apply $m(M,z)$,
while for merging (sub)halos, we compare the galaxy mass $m(M,z)$ to the sum 
over all its progenitors (in the previous timestep). If the galaxy mass from the model 
is higher than the sum of progenitor masses,
 we apply $m(M,z)$; if the galaxy mass is smaller, the new galaxy
mass is the sum over all its progenitors.

\section{Results}
\label{sec:results}

\subsection{Predicted halo mass functions from theory}
\label{halomftheory}
Before discussing the results of our numerical simulations, it is
worth presenting analytical predictions to gain insight into the
potential consequences of scale-dependent non-Gaussianities on
early structure formation. We here adopt a simple model and discuss
the effects on the HMF.

\begin{figure}
\centering
\includegraphics[width=0.7\hsize]{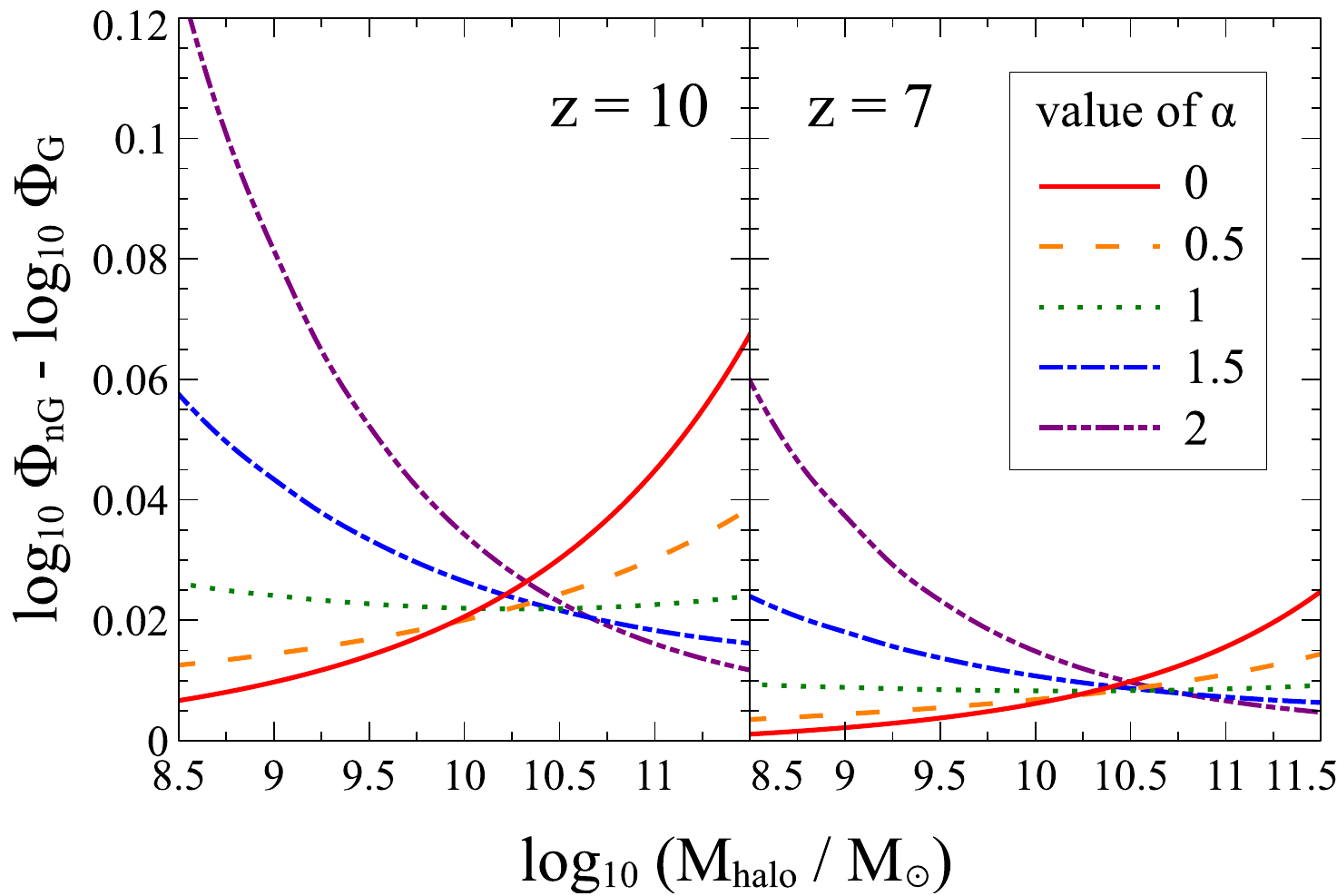}
\caption{Analytical predictions of the nG correction to the
  halo mass function. We plot the ratio of the halo mass function with
  non-Gaussian and Gaussian initial conditions at $z=10$ (left) and
  $z=7$ (right), for different running $f_{\rm NL}(k)$ passing through the
  pivot point of the first 3 models of Table~\ref{tab:fnlofk} (see Fig.~\ref{fig:fnlofk}).
\label{fig:analytical}}
\end{figure}

We follow the Press-Schechter formalism  \citep{Press74} for this calculation.
Namely, we work with the linear density field, $\delta_M$, smoothed with a spherical top
hat window that encompasses a mass $M$ and linearly extrapolated to
$z=0$, and consider that the one-point cumulants of this field uniquely
determine the HMF.  
Assuming that the nG correction is small, 
we apply the Edgeworth expansion to the one-point density probability
distribution function \citep{LoVerde+08}.
Up to the skewness order, the non-Gaussian to Gaussian ratio of the
HMF is given by
\begin{eqnarray}
\frac{\rmn{d}n_{\rm nG}/\rmn{d}M}{\rmn{d}n_{\rm G}/\rmn{d}M}(M,z) = 1 + \frac{1}{6}C_{M}^{(3)}H_3(\nu)
+ \frac{1}{6}\frac{{\rm d}C_{M}^{(3)}}{{\rm d}\ln\sigma_{M}}\frac{H_2(\nu)}{\nu},
\label{eq:MFanalytical}
\end{eqnarray}
where $\sigma^2(M)=\left\langle\delta_M^2\right\rangle$ is the variance
of the density fluctuations $\delta_M$, 
$C_M^{(3)}=\left\langle \delta_M^3\right\rangle / \sigma_M^3$
is a measure of the skewness of $\delta_M$,
$\nu = \delta_\mathrm{c}(z)/\sigma(M)$ is the peak height, given $\delta_\mathrm{c}(z) =
1.686/D_+(z)$, the threshold density
contrast for spherical collapse at redshift $z$,  where $D_+(z)$ is the growth rate,
and finally $H_n$ is the Hermite polynomial.

In this model, all the nG correction comes from the skewness,
which can be expressed by an integral of the {\it bispectrum}:
\begin{eqnarray}
\langle\delta_M^3\rangle = \int\frac{\mathrm{d}^3\mathbf{p}\mathrm{d}^3\mathbf{q}}{(2\pi)^6}\mathcal{M}(p)\mathcal{M}(q)\mathcal{M}(|\mathbf{p}+\mathbf{q}|)
B_\zeta(p,q,|\mathbf{p}+\mathbf{q}|),\label{eq:skew}
\end{eqnarray}
where $\mathcal{M}$ stands for the transfer function from the
curvature to the density fluctuation smoothed by a mass scale $M$, and
the bispectrum of the curvature $\zeta$ in the model (\ref{eq:localansatz}) is given by
\begin{eqnarray}
B_\zeta(k_1,k_2,k_3) = \frac{6}{5}\left[f_\mathrm{NL}(k_1)P_\zeta(k_2)P_\zeta(k_3) + ({\rm cyc.}\,2)\right],
\end{eqnarray}
where $({\rm cyc.}\,2)$ denotes two more terms that are obtained by
cyclic permutation of the wavenumbers 
in the first term.
The $k$ dependence of  $f_\mathrm{NL}$ propagates to the
mass dependence of skewness through these equations, making the
nG correction to the HMF rather non-trivial.
Since we focus on blue $f_\rmn{NL}$ (i.e., $\alpha>0$), we anticipate that
the correction to the HMF gets larger at low masses.

Fig.~\ref{fig:analytical} shows the analytical prediction~(eq.~[\ref{eq:MFanalytical}]) 
 at $z=10$ (left) and $z=7$ (right).
We here adopt $k_0 = 5.11\,h/\rm Mpc$, $f_{\rmn{NL},0}=18.5$, which is the intersection of the models 1, 2 and 3
(see Fig.~\ref{fig:fnlofk}), and vary the slope $\alpha$ as indicated in the figure legend. 
There are two noticeable trends in
Fig.~\ref{fig:analytical}. First, the dependence of the HMF ratio on $M$ depends on the slope
$\alpha$: the boost from non-Gaussianity is an increasing function of $M$ for 
$\alpha < 1$, while a larger $\alpha$ results in a decreasing
function of $M$. This high-mass enhancement of the HMF  for small $\alpha$ is
consistent with \citet{LoVerde+08}.
Because of the $\nu$-dependence in
equation~(\ref{eq:MFanalytical}), rare objects receive more
non-Gaussian effect in these cases.
When the $k$-dependence of $f_\rmn{NL}$ is blue enough,
it is the low-mass end of the HMF that 
is enhanced,
so that the mass dependence in $C_M^{(3)}$ overwhelms that of
$H_3(\nu)$.
Second, the nG correction is more prominent
at higher redshift. It is about a factor of two greater at $z=10$
compared to $z=7$. Although, not shown here, structure
formation at low redshift is almost unaffected with the models that
we consider here (i.e., the change of $\rmn{d}n/\rmn{d}M$ is less than
$10\%$ at $z<3$ over the mass range shown in Fig.~\ref{fig:analytical}).
Thus, early structure formation provides us with a unique opportunity to
constrain scale-dependent non-Gaussianity,
given the very tight \emph{Planck}
constraints on large scales.

\vspace {-0.5\baselineskip}

\subsection{Results from simulations}
\begin{figure*}
\includegraphics[width=0.6\hsize]{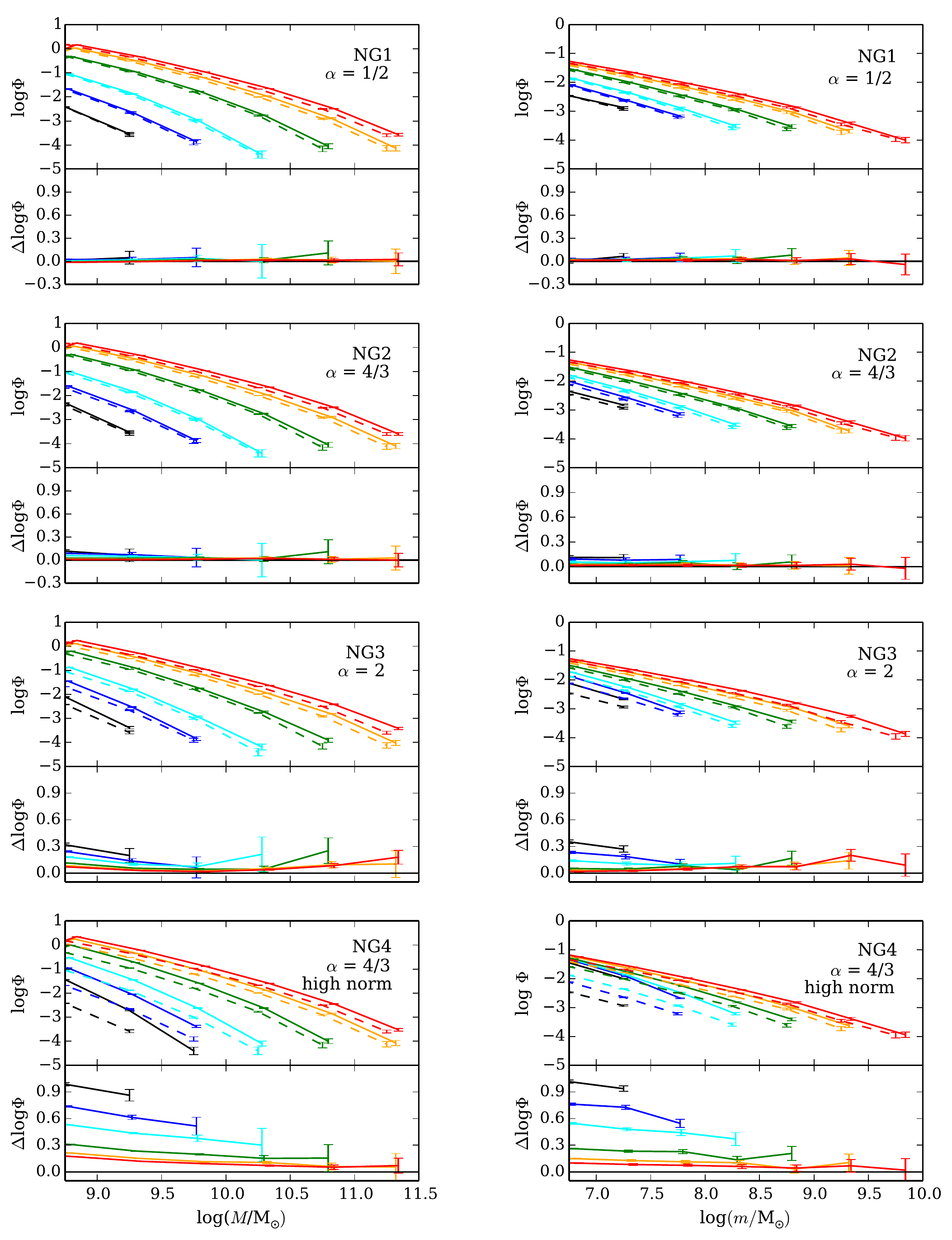}
\caption{\emph{Upper panels of boxes}: halo (\emph{left}) and galaxy stellar (\emph{right}) mass functions (in $\rm{Mpc}^{-3} \rm{dex}^{-1}$) for
  different initial conditions: Gaussian models (\emph{dashed}), and
  non-Gaussian models~1 to 4 (from \emph{top} to
  \emph{bottom}).
\emph{Lower panels of boxes}:  residuals of the log mass
function relative to that of the Gaussian run (the \emph{horizontal line}
  shows equal non-Gaussian and Gaussian mass functions). 
The errors are Poisson. The different curves indicate
different redshifts (decreasing upwards for the large boxes): 
$z=17$ (\emph{black}),
$z=15$ (\emph{blue}),
$z=13$ (\emph{cyan}),
$z=10$ (\emph{green}),
$z=8$ (\emph{orange}),
and
$z=7$ (\emph{red}).
\label{bigplot}}
\end{figure*}

The left-hand panels of Fig.~\ref{bigplot} show the HMFs
obtained from the five cosmological simulations.
One sees (upper left panel of Fig.~\ref{bigplot})
 that the effects of non-Gaussianity on
the HMF are increasingly important with increasing model number 
(see for example the upper left panel of Fig.~\ref{bigplot}).

Non-Gaussian
models~NG1 ($\alpha=1/2$) and NG2 ($\alpha=4/3$, low normalization) cause 
only small (less than 0.1 dex) and insignificant enhancements of the HMF 
(top two left panels of
Fig.~\ref{bigplot}) and SMF (top two right panels of Fig.~\ref{bigplot}).
Non-Gaussian model~NG3, with a very steep slope ($\alpha=2$), produces  significant enhancements (left panel in third row of
Fig.~\ref{bigplot})
of up to 0.3 dex ($z=17$) or 0.2 dex
($z=15$), in the HMF at   $\log M/\rmn{M_\odot} = 9$.
Finally, model~NG4, with a slope 4/3 but a much higher normalization (10 times that of model
NG2),
produces very large enhancements of the HMFs and SMFs at low masses and
high redshifts: greater than 0.3 dex enhancements in the HMF
arise for $z \geq 13$ at all halo masses and $z \geq 10$ for $\log
M/\rmn{M_\odot} < 9.5$.
The corresponding SMF is also enhanced by over 0.3 dex for
all galaxy masses at $z \geq 13$ and at galaxy  masses $\log m/\rmn{M_\odot}
\leq 6.8$ for $z=10$.

\vspace {-\baselineskip}

\section{Conclusions and discussion}
\label{sec:discussion}

The results presented here indicate that, in comparison with the predictions
from Gaussian ICs, simulations with ICs that are increasingly non-Gaussian at smaller scales,  yet
consistent with the CMB constraints from \emph{Planck},
can lead to small, but eventually detectable
alterations to the halo and galaxy stellar mass functions.
Since constant $f_{\rm NL}>0$ enhances
the HMF principally at large masses, one can think  that low slopes of
$\alpha={\rm d}\ln f_{\rm NL} / {\rm d}\ln k>0$ (keeping $f_{\rm NL} >0$) should also enhance the high-end of the HMF, while a high
enough slope should do the opposite and enhance the HMF at the low-mass end.
At $\alpha=2$, the HMF is in fact enhanced, both at the high
and low ends (left residual plot of third row of Fig.~\ref{bigplot},
although that of the high end
is only marginally
significant).
However, for the shallower slope ($\alpha=4/3$), the HMF is only
enhanced at the low end.
We find that our two strongest non-Gaussian models (NG3, NG4) exhibit the
largest differences,
 up to $0.2$ dex for NG3 ($\alpha=2$), and greater than $0.3$ dex (at $z=10$) for NG4 ($\alpha=4/3$, $f_{\rm NL,0}=10000$). 
These effects of nGICs on our simulated HMFs are close to the theoretical
predictions, with some quantitative differences.

Unfortunately, it is difficult to measure the HMF with great
accuracy, and considerably easier to measure the SMF.
We used the state-of-the-art model of stellar mass versus halo mass and
redshift of \citet{Behroozi+13} to produce galaxy masses on the (sub)halos of our cosmological $N$-body simulations. 
We slightly altered the model to
consider halo mergers and prevent galaxy masses from decreasing in time.
Comparing the resultant galaxy mass functions of our non-Gaussian models with
that of our Gaussian model, we find similar behavior of the
enhancements of the galaxy mass function with mass and redshift, i.e. $0.2$ dex for NG3, and $0.3$ dex (at $z=10$) for NG4.

The modification of the SMF by nGICs can have profound consequences.  In particular, the reionization of the Universe by the first stars and galaxies
will be affected, in a different way depending on the slope $\alpha$.  Low
mass galaxies are thought to be one of the most powerful sources of ionizing
photons at high redshift  \citep{Robertson+13,Wise2014}.
Using a set of cosmological  simulations, we
have seen that faint galaxies are the most affected by primordial
non-Gaussianities (see Models NG3 with $\alpha=2$ and 
NG4 with $\alpha=4/3$, but high normalization).  
Therefore primordial non-Gaussian perturbations can
then strongly affect the thermal history of the intergalactic medium. Effects
on the far-UV
luminosity function and the reionization history will be discussed in detail
in a forthcoming article (Chevallard et al., in prep).

\vspace {-\baselineskip}

\section*{Acknowledgments}
We are grateful to Doug Spolyar, Ben Wandelt, Herv\'e Aussel and Marta Volonteri for useful discussions.
TN is supported by Japan Society for the Promotion of Science (JSPS)
Postdoctoral Fellowships for Research Abroad. 

\vspace {-\baselineskip}

\bibliography{biblio}

\label{lastpage}

\end{document}